# Resonant Raman scattering of surface phonon polaritons mediated by excitons in WSe$_2$ films

**Lanqing Zhou[1,2], Konstantin Wirth[3], Minh N. Bui[1,2], Renu Rani[1,2], Detlev Grützmacher[1,2], Thomas Taubner[3] and Beata E. Kardynał[1,2]\***

[1] Peter Grünberg Institute 9, Forschungszentrum Jülich, D-52425 Jülich, Germany

[2] Department of Physics, RWTH Aachen University, D-52074 Aachen, Germany

[3] 1st Institute of Physics (IA), RWTH Aachen University, D-52074 Aachen, Germany

## Abstract

Surface phonon-polaritons propagating along interfaces of polar dielectrics coexist with excitons in many van der Waals heterostructures, so understanding their mutual interactions is of great interest. Here, we investigate the type I surface phonon polariton of hBN via low-temperature resonant-Raman spectroscopy in hBN/WSe$_2$ heterostructures. The resonantly enhanced hBN surface phonon polariton (SPhP) Raman signal, when laser energy is such that the scattered photons have energy close to that of the WSe$_2$ excitons, enables detailed characterization of type I SPhP in hBN even when hBN is one monolayer thick. We find that the measured bandwidth of the SPhP Raman signal depends on the thicknesses of the hBN layer. We are able explain the experimental data using transfer matrix method simulations of SPhP dispersions providing that we assume the Raman scattering to be momentum non-conserving, as could be the case if localized WSe$_2$ exciton states participated in the process. We further show that resonant Raman scattering from SiO$_2$ SPhP can also be mediated by WSe$_2$.

Polar, uniaxial, hexagonal boron nitride (hBN) features two Reststrahlen bands due to the degeneracy breaking of longitudinal optical (LO) and transverse optical (TO) phonon frequencies. Within each Reststrahlen band, the basal component of the permittivity tensor ($\varepsilon_\parallel$) and the one along the optic axis ($\varepsilon_\perp$) have opposite signs, with a lower band ($\varepsilon_\parallel>0$, $\varepsilon_\perp<0$) at around 800 cm$^{-1}$, referred to as type I hyperbolic band, and the higher band ($\varepsilon_\parallel<0$, $\varepsilon_\perp>0$ ) at about 1350 cm$^{-1}$ called type II hyperbolic band [1-6]. It has been shown that surface phonon polaritons (SPhPs) of hBN within these bands can be used to confine and manipulate light at deep subwavelength scale, used for sub-diffraction imaging and surface-enhanced infrared spectroscopy [7-10]. The hBN SPhPs are similar with surface plasmon polaritons but exhibit much lower optical losses [7, 11, 12].

Methods based on scattering-type scanning near-field optical microscopy(s-SNOM), offering high-resolution imaging SPhP in real space, are widely used to probe type II hBN SPhP [1, 13, 14]. In addition, the van der Waals bonding of the hBN monolayers facilitates the control of both the wavelength and the amplitude of polaritonic waves with the thickness of hBN films [15]. Highly confined type II SPhPs were experimentally observed even in the monolayer limit, with the LO and TO phonon modes degenerate at $\Gamma$ point [16, 17]. In contrast, the excitation of lower energy band, type I hBN SPhPs with s-SNOM were barely seen, especially for thin layers due to weak coupling between the tip and the strongly confined electric field inside hBN for out-of-plane polariton [18, 19].

\* b.kardynal@fz-juelich.de

Resonant Raman sensing can be used as an alternative since it is a powerful method to measure weak photon scattering [20, 21] especially as they are performed at near-infrared wavelength range using standard light sources and detectors. In addition, the method may provide direct access to large momenta of surface phonon polaritons without near-field requirements or other wavevector matching strategies [22, 23]. However, it requires the right material system with a strong optical transition in the excitation or detection energy range. Excitonic states of transition metal dichalcogenides (TMDs) have been shown to facilitate resonant Raman sensing of type I hBN SPhPs [24-29]. For example, extreme enhancement of the Raman signal from surface phonon polariton scattering in the thick hBN/monolayer $WSe_2$ has been obtained with the energy of incoming photons matching the 2s excitonic state of $WSe_2$ and with the energy of outgoing photons close to that of 1s excitonic state.

In this work, we use a multilayer $WSe_2$ as a sensor in resonant Raman scattering to study type I surface phonon polaritons that emerge in hBN layers of variable thickness interfaced with $WSe_2$. We choose indirect-bandgap multilayer $WSe_2$ to avoid photoluminescence mixing with the Raman signal. We measure hBN layers with thicknesses from a monolayer (1L) to a bulk-like to get insights into the evolution of the polariton bandwidth. We extract the experimentally obtained Raman bandwidths and compare them with our calculations of the hBN phonon-polariton bandwidth to find that we need a momentum-nonconserving process to explain our results.

We investigated hBN/$WSe_2$ heterostructures supported on a 300 nm thick $SiO_2$ on Si substrate. The fabrication of the samples started from an exfoliation of $WSe_2$ and hBN flakes (from HQ graphene) from bulk crystals onto $SiO_2$/Si substrates. Subsequently, the flakes were picked up with a polycarbonate stamp one by one to assemble the heterostructure, and the complete multilayer stack was then released on the target substrate [30]. During sample preparation, the interface between $WSe_2$ and hBN had no contact with polymers. Different heterostructure geometries were studied, with $WSe_2$ or hBN in contact with $SiO_2$ and one where $WSe_2$ was sandwiched between the hBN flakes.

The micrograph in Figure 1(a) shows a typical sample with hBN on top of $WSe_2$, on a $SiO_2$/Si substrate. The hBN flake contains areas of a different number of monolayers, whose thickness has been determined from an atomic force microscopy (AFM) image shown in Figure 1(b). The inset in Figure 1(b) shows a height profile of the hBN flake along the black arrow. Since the thickness of 1L hBN is around 0.5 nm, the areas of different heights are 1L, 3L and 5L thick. The apparent thickness of the monolayer of hBN on $SiO_2$ is approximately 1 nm, possibly enlarged by the air gap between hBN and $SiO_2$ substrate. To avoid ambiguities, the layer numbers of hBN used in our experiments is further confirmed by room temperature Raman spectroscopy [31], results of which can be seen in Figure S1 in Supplemental Material [32].

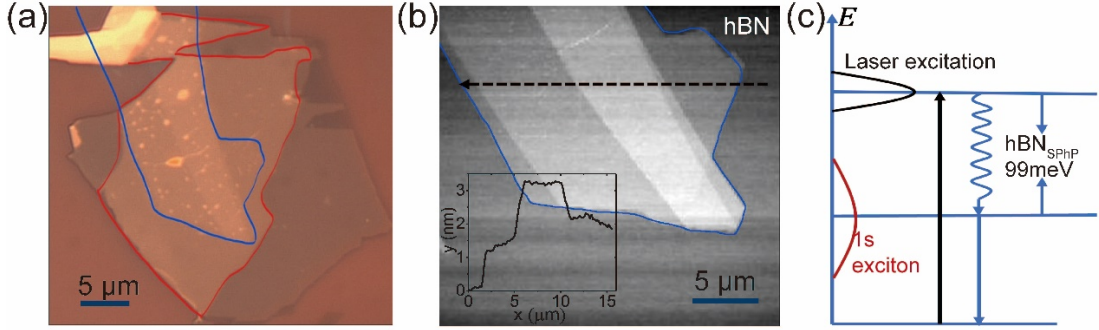

Figure 1. (a) An optical micrograph of the sample. Four-layer WSe$_2$ outlined in red is partly covered by an hBN flake, outlined in blue. The hBN flake contains areas of different thicknesses, which can be seen in the AFM image of the hBN flake on SiO$_2$ after exfoliation and before stacking with WSe$_2$ shown in (b). The height profile along the black arrow in shown in the inset in (b). (c) Schematic diagram illustrating resonant Raman scattering process under consideration in this work. The heterostructure is excited with a laser which is blue-detuned (black arrow) relatively to the 1s exciton of WSe$_2$. When the detuning is close to the energy of type I hBN SPhP (99 meV), a strong Raman signal (blue, straight arrow) appears in the spectra.

We performed Raman scattering measurements on the heterostructures at 10 K while varying the laser excitation energy in resonant-on channel, which means that hBN SPhP the scattered photons energetically matched WSe$_2$ 1s exciton emission energy, as schematically shown in Figure 1(c). The excitation source was a continuous wave tunable Ti:sapphire laser. The collimated laser beam, normal to the hBN film, was focused on the sample using a lens with numerical aperture (NA) of 0.47. The scattered photons were collected by the same lens and refocused on the entrance slit of the Turner-Czerny spectrometer. Figure 2 shows Raman spectra from 16L hBN/4L WSe$_2$ sample obtained with different laser energies from 1.7971 eV to 1.8535 eV. The spectra show broad lines centered at 780 cm$^{-1}$, originating from scattering with type I hBN surface phonon polaritons confined within the Reststrahlen band. The detected signal is the strongest and symmetric for a laser energy of around 1.837 eV corresponding to 1.738 eV of detection energy. When the laser is slightly blue- or red-detuned from this energy, the signal becomes asymmetric as the laser is resonant with the upper or lower side of the SPhP band. At large detuning the signal becomes weak. This evolution of the Raman signal is consistent with the 1s exciton resonance at 1.738 eV, which can be seen in Figure S2 in Supplemental Material [32] showing the data in energy units.

The evolution of the Raman signal can be compared with the density of states arising from the dispersion of the surface phonon polaritons [6, 15, 19, 33]. We performed phonon polariton dispersion simulations using a transfer matrix method (TMM) [34]. In the simulations, we used the layer structure of our samples and dielectric function from reference [35] for hBN and reference [36] for SiO$_2$. The static dielectric constants of 15 and 15.8, were used for WSe$_2$ and Si, respectively [37]. Figure 3(a) and (b) show the simulated dispersions relations of SPhP at the interface of 4L WSe$_2$ below 16L and 400L thick hBN layer, respectively. The bulk-like 400L hBN was chosen to illustrate thickness dependence of the dispersion [15, 33]. The poles of the imaginary part of Fresnel reflection coefficients, indicated by magenta dashed curves in Figure 3, show the SPhP modes of hBN film.

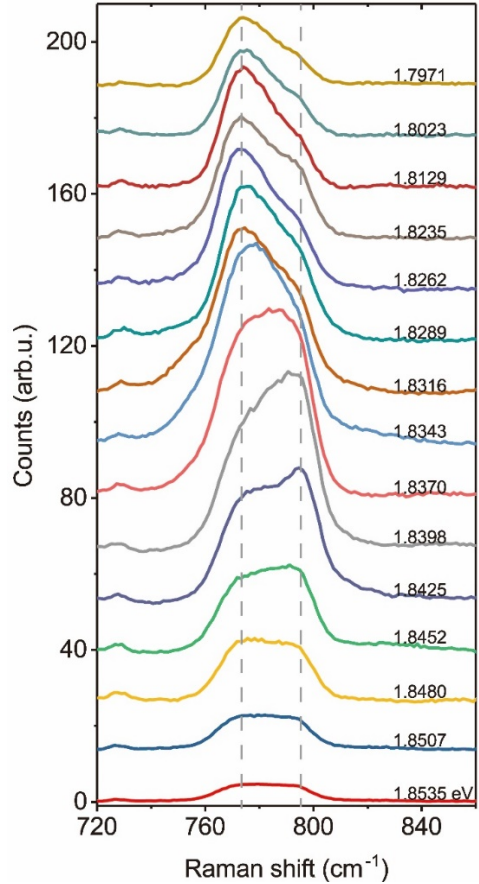

Figure 2. Raman spectra collected from the sample with a 16L hBN flake covering the 4L WSe$_2$. Each spectrum is obtained on excitation with a different laser energy, given in the label. The observed line shape depends on the laser's energy relative to the resonance's peak. The strongest resonance occurs when the laser energy is 1.837 eV and the detection energy 1.738eV, which matches WSe$_2$ 1s exciton energy. When the laser is blue- or red-detuned from 1.837 eV, the signal becomes asymmetric with less intensity. The gray lines are guides for the eye.

In our experiments, the maximum angle of incidence of excitation photons is given by the 0.47 NA of the objective lens. This means that the maximum in-plane momentum in the momentum conserving processes is $4.4\times10^4$ cm$^{-1}$. We then find the expected energy range of the Raman signal for 16L hBN, by integrating the maximum imaginary part of Fresnel reflection coefficients with k-vectors up to this critical value, marked by a vertical orange line in Figure 3(a). The resulting line, shown in blue in Figure 3(c), extends in a narrow band from 812 to 819 cm$^{-1}$ with bandwidth 7 cm$^{-1}$. This is an upper limit for the line width as it ignores energy dependent matrix element. Meanwhile, the experimental bandwidth from the sample, in Figure 2, is dramatically larger with the full width at half maximum (FWHM) of 46 cm$^{-1}$ at laser energy of 1.837 eV, even when factoring in the system response broadening of 5.8 cm$^{-1}$ (Figure S3 in Supplemental Material [32]). The measured line extents over the full energy range of the simulated SPhPs, including states at very high k-vectors ($2.0\times10^6$ cm$^{-1}$), as shown in the green line in Figure 3(c). This means that a wider range of laser energies satisfies the resonant Raman condition. Particularly, it can be tuned by the width of the SPhPs dispersion. The density of states is very high near the band edges and if the laser energy fulfils the resonance condition at the band edge energies, there is a strong enhancement of the signal as we see in Figure 2. The low frequency edge would be inaccessible in a k-conserving case.

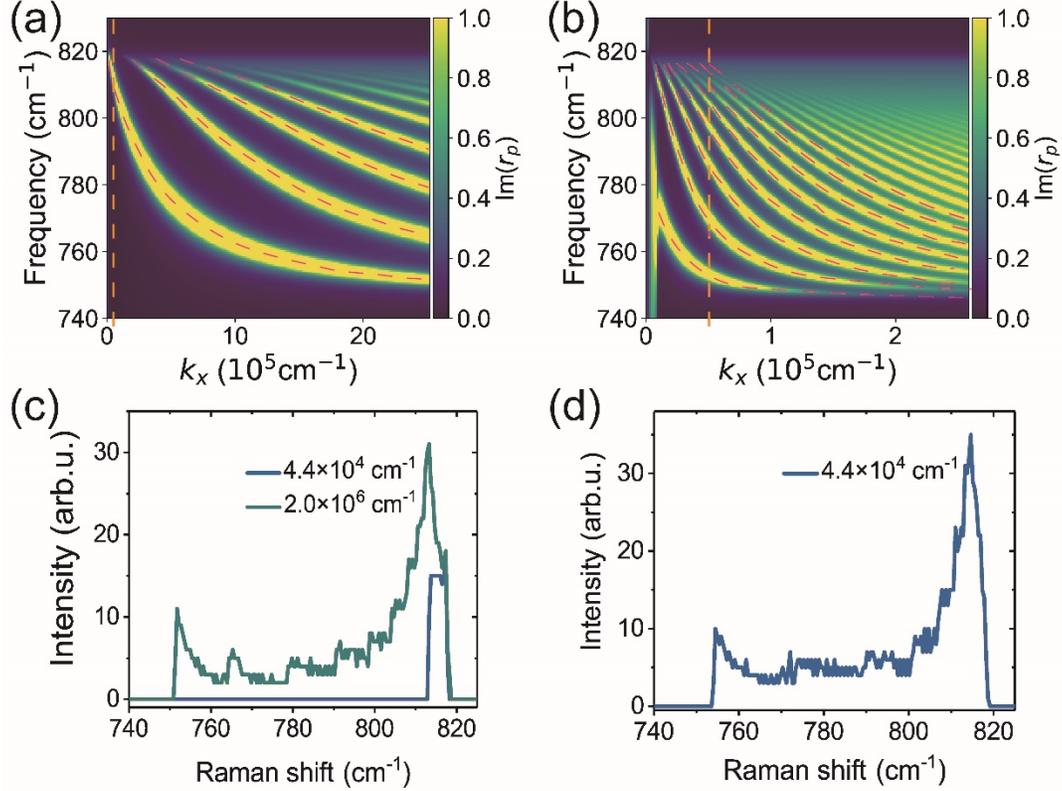

Figure 3. Contour plot of simulated dispersion relation for surface phonon polaritons at the interface of 4L WSe$_2$ below (a) 16L and (b) 400L thick hBN layer on SiO$_2$ with z-axis plotting the imaginary part of Fresnel reflection coefficient, SPhP modes are manifested by a maximum reflection (indicated by magenta dashed curves). The maximum k-vector that can be reached with our NA is marked with a vertical dashed orange line, as shown in (a)(b). (c)(d) The spectra obtained by an integration of the maximum imaginary part of Fresnel reflection coefficient up to the maximum k-vector $4.4 \times 10^4$ cm$^{-1}$, corresponding to (a) and (b), see blue curves. The green curve in (c) is the integrated spectrum with higher kvector at $2.0 \times 10^6$ cm$^{-1}$.

As seen in Figure 3(c) and (d), the width of the SPhPs band depends on the thickness of the hBN layer. To get further insight into the scattering process and the origin of the large width of SPhP signal, we measured samples that contain hBN layers with different thicknesses at fixed excitation laser energy of 1.837 eV to remain at resonance. Figure 4(a) shows Raman spectra of SPhP measured in heterostructure that contained thicker hBN layers 25L, 46L and 66L. Notably, the three spectra are almost identical to each other. It is not the case for the Raman spectra of the samples with thinner hBN layers, down to a single monolayer, which are shown in Figure 4(b). When decreasing the hBN layer thickness from ten to a single monolayer, the Raman signal of SPhP red-shifts and becomes narrower.

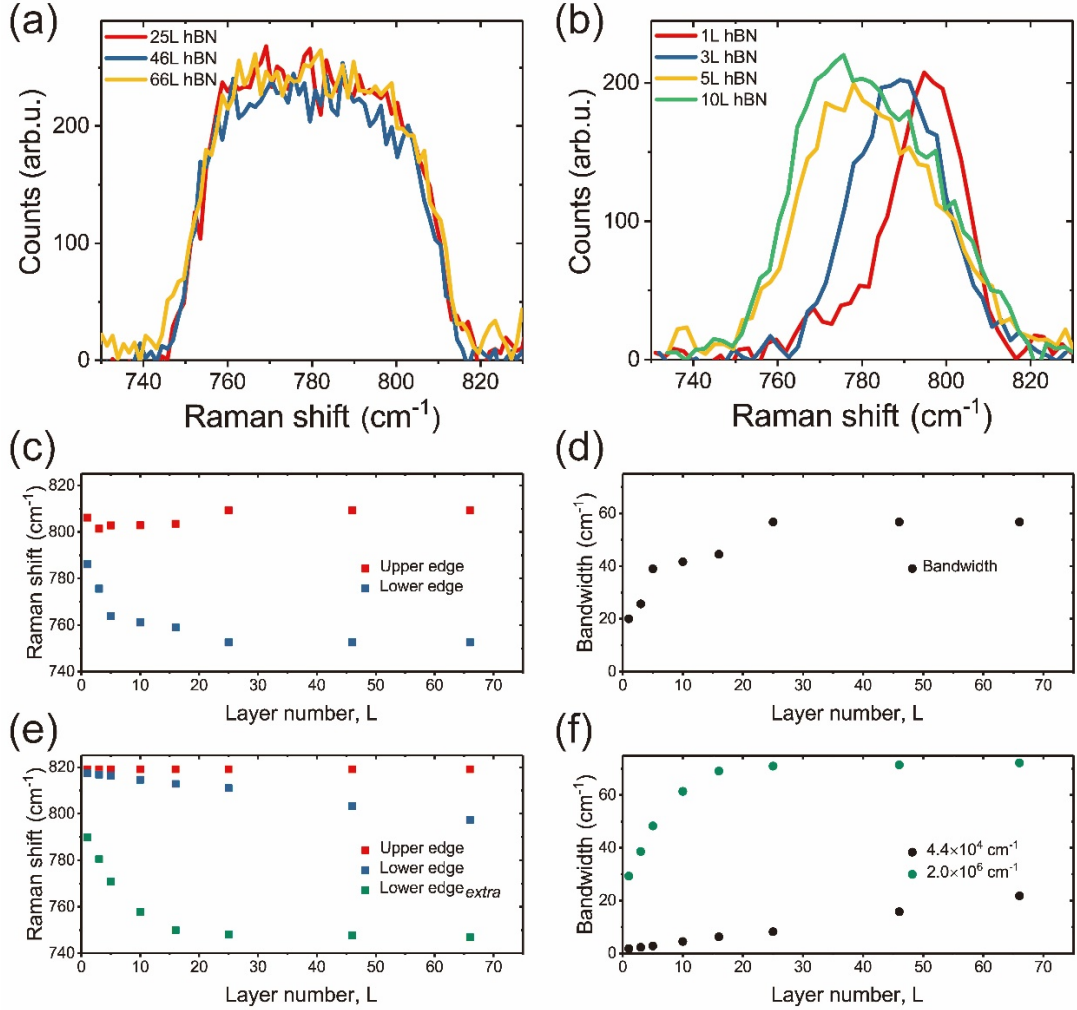

Figure 4. Profiles of the hBN Raman spectra originating from type I surface phonon polaritons for different layers of hBN (a) 66L, 46L and 25L, (b) 10L, 5L, 3L and 1L. (c) Upper band edge (red) and lower band edge (blue) of the experimentally obtained Raman spectra as a function of hBN layer number L and (d) the corresponding bandwidth defined as difference between the upper and lower band edges. (e) Upper band edge (red) and lower band edge (blue) got from the simulation as a function of hBN layer number L and (f) the corresponding bandwidth with k vector at $4.4\times10^4$ cm$^{-1}$. The green symbol in (e) represents simulated lower edge with larger k vector at $2.0\times10^6$ cm$^{-1}$ and the corresponding bandwidth in (f), as the green symbol illustrates.

To have a better comparison between the experimental and simulated data, we plot the thickness dependence of the experimental lower band edge (blue) and upper band edge (red), defined as a half of the line amplitude in Figure 4(a) and (b), in Figure 4(c). The corresponding bandwidth which is defined as the difference between upper and lower band edges is plotted in Figure 4(d). With the choice of maximum momentum cut-off at $4.4\times10^4$ cm$^{-1}$ dictated by NA of our lens for a momentum conserving process, the band edges and bandwidths extracted from the simulations (shown in Figure S4 in Supplemental Material [32]) are plotted in Figure 4(e) and Figure 4(f), respectively. In both Figures 4(c) and (e), the position of upper band edge does not change with the hBN thickness. The experimental and calculated lower band edge frequency both decrease with layer numbers due to modifications of interactions between atoms [15, 17]. While the simulated lower band edge appears at much higher value of Raman shift and consequently a very narrow bandwidth (black in Figure 4(f)). In order to match simulations with the experimental data, we extended the maximum wavevectors of the polariton that can be excited in the scattering process. When the maximum allowed wavevector is set to $2.0\times10^6$ cm$^{-1}$, the extracted lower band edge (Lower edge$_{extra}$) lies at lower value of Raman shift and therefore broadened bandwidth, as also shown in Figure 4(e) and (f), respectively. The agreement between the measured and simulated lower band edge and therefore bandwidth is now reasonable for all hBN thicknesses, including the monolayer. The simulated bandwidth for the thicker layers exceeds, however, the measured one and the match could be improved by selecting smaller cut-off k-vectors. It is likely that the widths of the measured SPhPs bands for the thinnest layers are broadened due to shorter lifetimes of the excitations compared with the thicker layers and that smaller cut-off k-vectors would also be more appropriate. While deconvolution of the linewidth broadening and therefore more accurate cutoff momentum is beyond the scope of this study, the simulations confirm that scattering from the hBN SPhP with wavevectors larger than allowed by momentum conservation contribute to the measured signal.

The finding that Raman scattering involves excitation of SPhPs at high k-vectors is supported by a theoretical study which showed that SPhP in hBN in several layers limit requires momentum transfer above $10^6$ cm$^{-1}$ [38]. Furthermore, type II hBN SPhPs in monolayer h-BN and in 10-nm-thick h-BN flakes were found to extend over nearly the whole Reststrahlen band when measured with electron energy-loss spectroscopy which can achieve high momentum transfer above $10^6$ cm$^{-1}$ [16, 39]. In our experiments, the momentum transfer can be possible if excitons participating in the resonant Raman process are localized in even smallest potential fluctuations. In such a case, excitonic wavefunction is delocalized in the k-space, and contains a range of momenta that can be transferred to the SPhPs [40, 41].

Further, we investigated samples with geometries that lead to potentially different overlap of hBN SPhPs with SiO$_2$ [42-44]. We prepared samples that integrate heterostructure with different configurations: (i) WSe$_2$-covered hBN (hBN/WSe$_2$/SiO$_2$), (ii) hBN-covered WSe$_2$ (WSe$_2$/hBN/SiO$_2$) and (iii) hBN-encapsulated WSe$_2$ (hBN/WSe$_2$/hBN/SiO$_2$), see Figure S5 in Supplemental Material [32] for the optical images of samples geometries. The spectra from these samples are collected in Figure 5(a). Apart from the discussed above type I SPhPs from hBN, there are two lines in the figure. The peak centered at 1020 cm$^{-1}$ present in all samples is WSe$_2$ phonon line caused by exciton energy loss [45-48]. The line at 1110 cm$^{-1}$, which we label with a red asterisk, is observed only in the sample with WSe$_2$ in direct contact with SiO$_2$ (sample (i)) and is the surface phonon polariton of SiO$_2$. It is also absent in the Raman spectra of SiO$_2$ (see Figure S6 in Supplemental Material [32]) measured under the same excitation. In Figure 5(b), we plot the extracted intensity of the SiO$_2$ SPhP Raman signal as a function of laser energy, the resonant behavior can be observed. Its presence indicates that WSe$_2$ makes sufficient contact with non-layered materials as well as hBN [25, 49-53] and the excitation of SiO$_2$ SPhP enlarges the family of dielectrics whose SPhP can be excited, which is mediated by excitons states of 2D semiconductor.

Looking at hBN SPhP peak in the three spectra, there is not much difference in hBN SPhP bandwidth among the three stacks while the intensity varies. The intensity from hBN-covered WSe$_2$ sample is higher than that of WSe$_2$-covered hBN sample. The intensity from hBN-encapsulated WSe$_2$ sample is more than two times higher than that from hBN-covered WSe$_2$ sample. The difference between the samples with different geometries may be related to the dependence of interfacial excited SPhP or of the properties of excitons in WSe$_2$ in heterostructures on the sample environment and its manipulation [36, 54]. It is unclear whether SiO$_2$ affects the intensity of hBN SPhP by, for example, changing their lifetime by scattering on its surface. The scattering could also be related to the SPhP launched in SiO$_2$ discussed above.

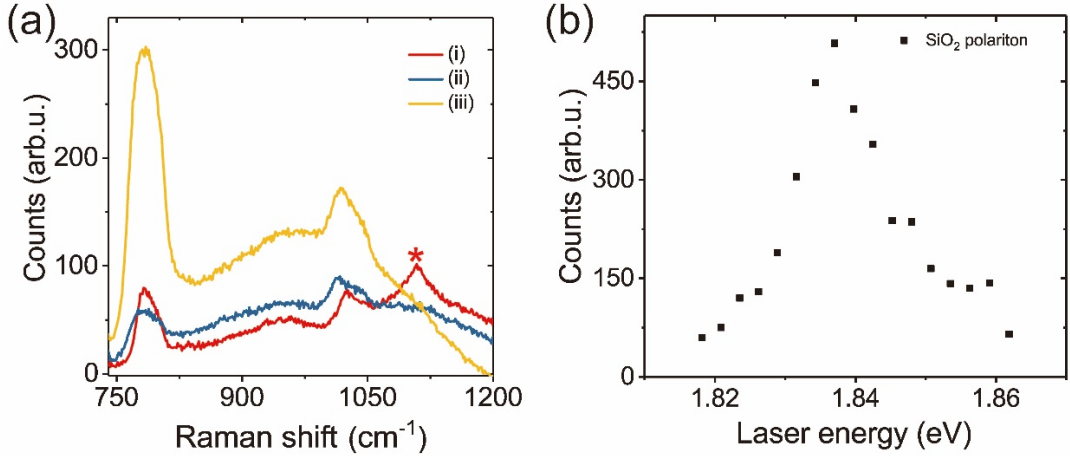

Figure 5. (a) Raman spectra collected from the sample with different configurations:(i) hBN-covered WSe$_2$ (hBN/WSe$_2$/SiO$_2$), (ii) WSe$_2$-covered hBN (WSe$_2$/hBN/SiO$_2$) and (iii) hBN-encapsulated WSe$_2$ (hBN/WSe$_2$/hBN/SiO$_2$). The SiO$_2$ SPhP peak labelled with the red asterisk was collected from hBN-covered WSe$_2$ region. (b) Integrated Raman intensity of SiO$_2$ SPhP as a function of laser energy, showing resonant character.

In summary, we have conducted an experimental investigation of type I phonon polaritons' evolution on the interface of hBN with WSe$_2$ via resonant Raman scattering at low temperatures, which is not easily accessible for a few-layer films with other method. We have found that the Raman line was present even when WSe$_2$ is in contact with a monolayer thick hBN. The spectra excited with lasers at different energies show the same line width but greatly modulated line shape. Simulations of the SPhPs dispersion in films of the same thickness consistently predict narrower line widths than the measured ones when we assume momentum-conserving scattering. A good match is obtained if we allow for large wavevector transfer to the SPhPs, which could be facilitated by localized WSe$_2$ excitons. The environmental effects in our measurements of samples with different geometries suggest that careful design of samples may lead to even larger signal enhancement and new insights into SPhPs. It would, for example, be interesting to transfer the heterostructures to different substrates or suspend them for measurements. Prospectively, it should be possible with much higher exciton generation rate to fulfill the condition of hyperbolic phonon polaritons in WSe$_2$ at similar frequencies to those of hBN and which depends on exciton density. Such heterostructure could facilitate investigation of resonant interactions between different polaritons. We have also shown the resonant Raman enhancement of surface phonon polaritons of SiO$_2$ in contact with WSe$_2$ suggesting that WSe$_2$ could be used to probe other dielectrics that support surface polariton at similar energy ranges, e.g., GaN and $MoO_3$.

# Supporting information

**Lanqing Zhou[1,2], Konstantin Wirth[3], Minh N. Bui[1,2], Renu Rani[1,2], Detlev Grützmacher[1,2], Thomas Taubner[3] and Beata E. Kardynał[1,2]**\*

[1] Peter Grünberg Institute 9, Forschungszentrum Jülich, D-52425 Jülich, Germany
[2] Department of Physics, RWTH Aachen University, D-52074 Aachen, Germany
[3] 1st Institute of Physics (IA), RWTH Aachen University, D-52074 Aachen, Germany

\* b.kardynal@fz-juelich.de

1. **Room temperature Raman spectroscopy for hBN thickness assessment.**

Figure S1 displays the $E_{2g}$ Raman line obtained from sample in Figure 1. The signal originating from the monolayer area is centered at 1369 cm$^{-1}$ and exhibits a redshift of 3 cm$^{-1}$ compared to the $E_{2g}$ Raman line of the bulk material. The shift, attributed to a slightly shorter B–N bond expected in an isolated monolayer [31], serves as a confirmation that the thinnest region corresponds to a monolayer. The intensities of $E_{2g}$ Raman lines, as illustrated in Figure S1 for various regions of the hBN flake, exhibit a proportional scaling with different hBN thicknesses when the system is calibrated [31]. This scaling aligns well with the determination obtained from the AFM scan.

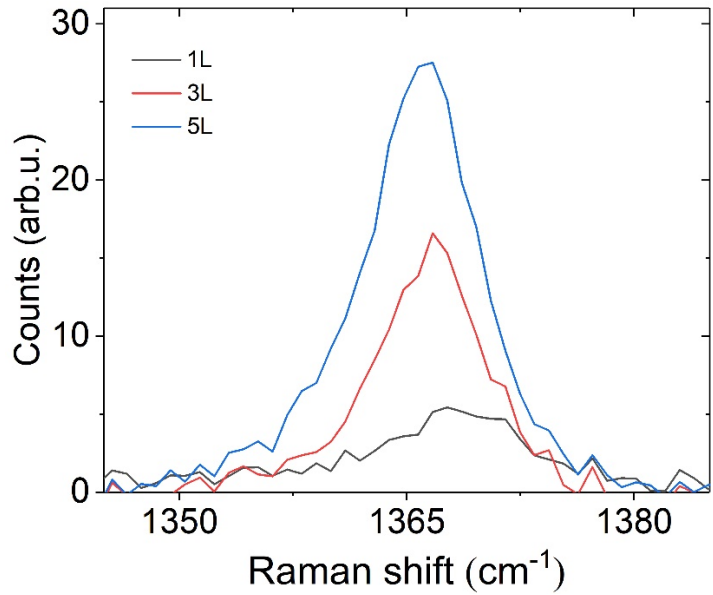

Figure S1. $E_{2g}$ Raman line of a monolayer, trilayer and five-layer hBN measured at room temperature. The Raman line shift and its intensity increases proportionally with the number of layers.

## 2. Shape variation of hBN SPhP on both sides of WSe$_2$ exciton emission resonance.

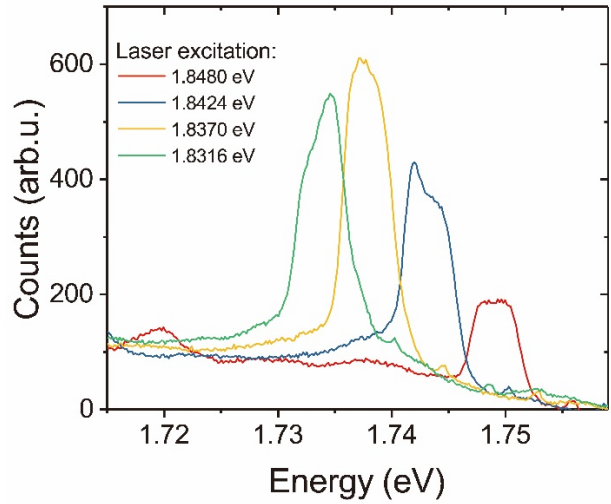

Figure S2. Spectra from Figure 2 plotted as a function of laser energy. Notably, a portion of the Raman peak exhibits increased sensitivity to laser energy, particularly when the Raman line is in proximity to the A exciton position at 1.738eV.

**3.    Low temperature Silicon Raman line.**

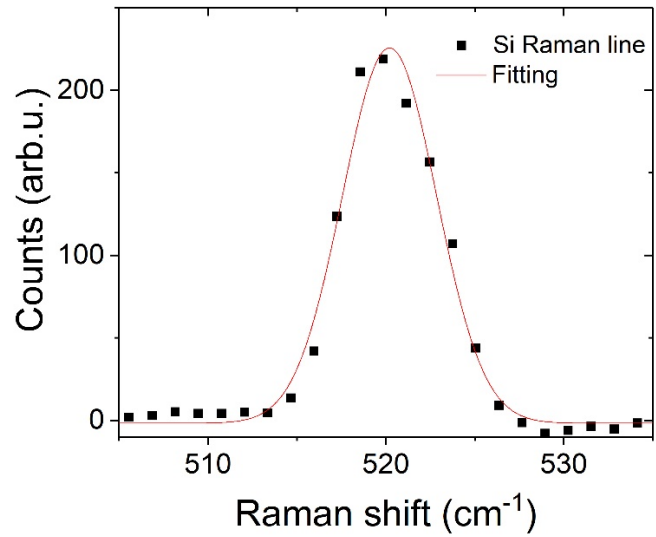

Figure S3. Raman line at 520 cm$^{-1}$ from Si obtained using the same setup employed for polariton sensing. The FWHM of this line with Voigt profile, measured at 5.8 cm$^{-1}$, represents the upper limit for instrumental broadening in our measurements.

4. **Simulation: varying thickness effects on surface phonon polariton dispersion in hBN.**

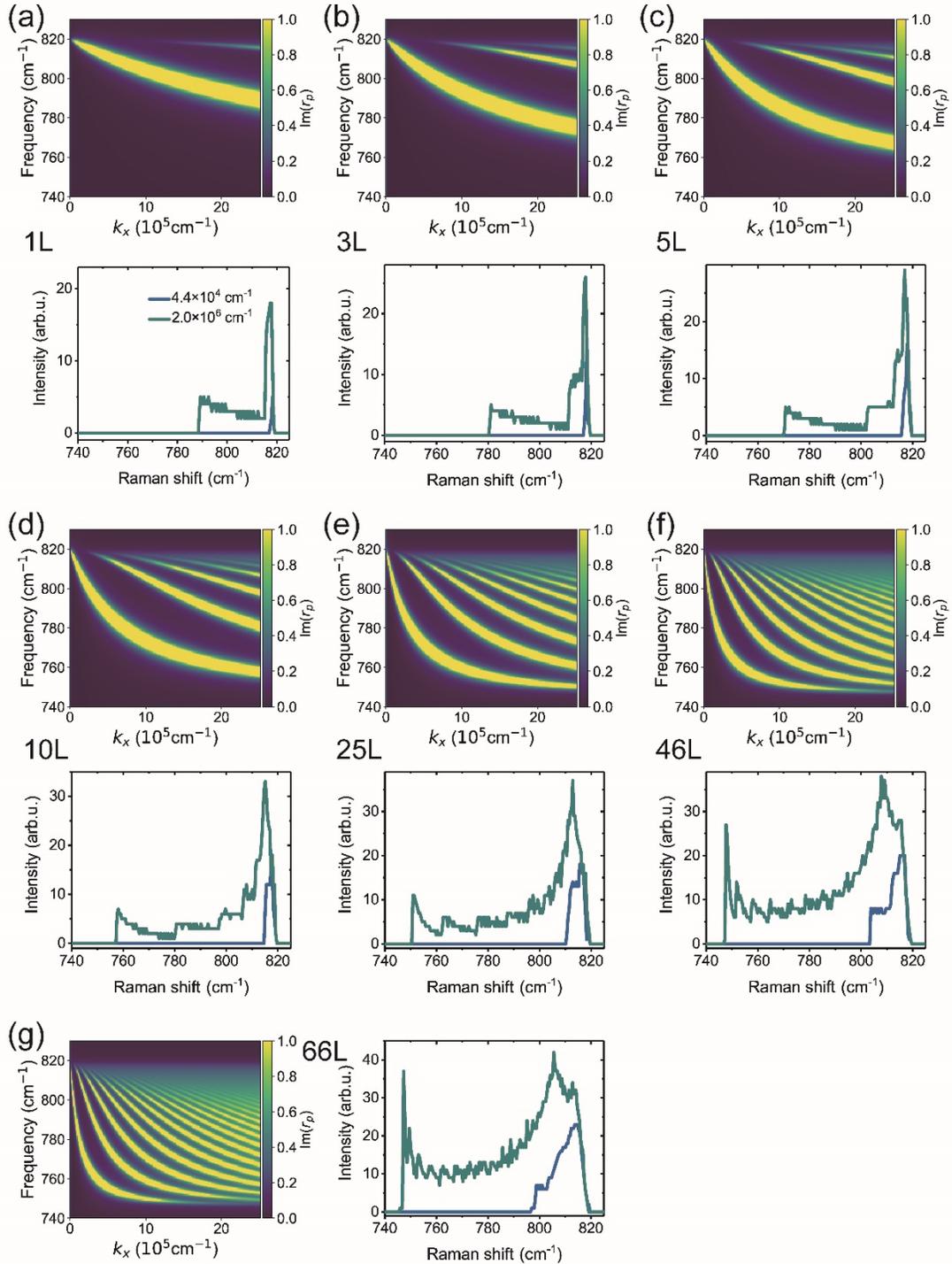

Figure S4. Simulation of surface phonon polariton dispersion of hBN with thickness 1L (a), 3L (b), 5L (c),10L (d), 16L (e), 25L (f), 46L (g) and 66L (h), with the corresponding spectra expected from the integration of imaginary part of Fresnel reflection coefficient over all k vectors until maximum k at $0.44\times10^5$ cm$^{-1}$(blue line) and $2.0\times10^6$ cm$^{-1}$(green

line).

**5. Description of sample geometry in Figure 4.**

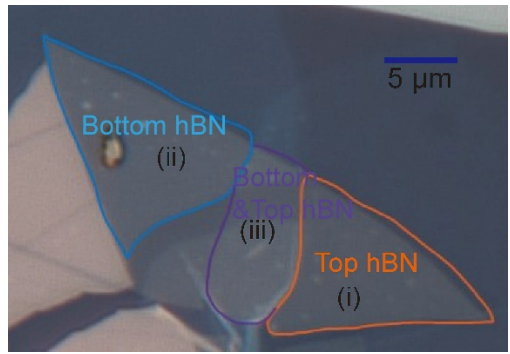

Figure S5. Various sample geometries. The sample consists of different geometries. (i) hBN-covered $WSe_2$ (hBN/$WSe_2$/$SiO_2$), (ii) $WSe_2$-covered hBN ($WSe_2$/hBN/$SiO_2$) and (iii) hBN-encapsulated $WSe_2$ (hBN/$WSe_2$/hBN/$SiO_2$). The thickness of the top hBN is 16L, and the bottom one is 31L.

## 6. Spectra of SiO$_2$ Surface phonon polariton obtained from WSe$_2$/ SiO$_2$ sample.

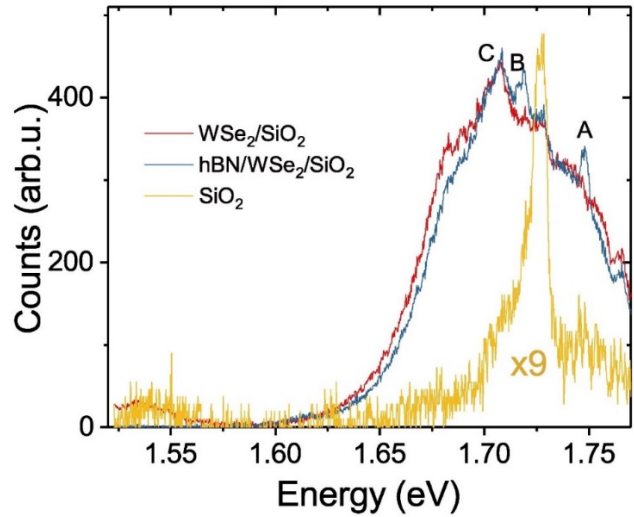

Figure S6. Surface phonon polariton of SiO$_2$ in WSe$_2$/SiO$_2$ region under laser excitation at 1.84 eV. Raman modes A (hBN SPhP), B (WSe$_2$ phonon) and C (SiO$_2$ SPhP) appear in the hBN/WSe$_2$/SiO$_2$ area. Only mode C appears in WSe$_2$/SiO$_2$ area. In pristine SiO$_2$ substrate, we only observe a broad Raman feature, attributed to the second-order Raman of Si. This graph further confirms the brightening of SiO$_2$ surface phonon polariton in contact with WSe$_2$.